\documentclass[
showpacs,preprintnumbers,amsmath,amssymb]{revtex4}
\usepackage{amsmath}
\usepackage{graphicx}
\usepackage{subfig}
\usepackage{hyperref}
\usepackage{amssymb}
\usepackage[english]{babel}
\usepackage{epsfig}
\usepackage{wasysym}
\usepackage{graphicx}
\usepackage{indentfirst}
\usepackage{amsmath}
\usepackage{amsfonts}
\usepackage{amssymb}
\usepackage{enumerate}

\begin{document}

\newcommand{\vecbo}[1]{\mbox{\boldmath $#1$}}
\newtheorem{theorem}{Theorem}
\newtheorem{acknowledgement}[theorem]{Acknowledgement}
\newtheorem{algorithm}[theorem]{Algorithm}
\newtheorem{axiom}[theorem]{Axiom}
\newtheorem{claim}[theorem]{Claim}
\newtheorem{conclusion}[theorem]{Conclusion}
\newtheorem{condition}[theorem]{Condition}
\newtheorem{conjecture}[theorem]{Conjecture}
\newtheorem{corollary}[theorem]{Corollary}
\newtheorem{criterion}[theorem]{Criterion}
\newtheorem{definition}[theorem]{Definition}
\newtheorem{example}[theorem]{Example}
\newtheorem{exercise}[theorem]{Exercise}
\newtheorem{lemma}[theorem]{Lemma}
\newtheorem{notation}[theorem]{Notation}
\newtheorem{problem}[theorem]{Problem}
\newtheorem{proposition}[theorem]{Proposition}
\newtheorem{remark}[theorem]{Remark}
\newtheorem{solution}[theorem]{Solution}
\newtheorem{summary}[theorem]{Summary}
\newenvironment{proof}[1][Proof]{\textbf{#1.} }{\ \rule{0.5em}{0.5em}}
\hypersetup{colorlinks,citecolor=green,filecolor=magenta,linkcolor=red,urlcolor=cyan,pdftex}

\newcommand{\be}{\begin{equation}}
\newcommand{\ee}{\end{equation}}
\newcommand{\bea}{\begin{eqnarray}}
\newcommand{\eea}{\end{eqnarray}}
\newcommand{\beaa}{\begin{eqnarray*}}
\newcommand{\eeaa}{\end{eqnarray*}}
\newcommand{\Lhat}{\widehat{\mathcal{L}}}
\newcommand{\nn}{\nonumber \\}
\newcommand{\e}{{\rm e}}


\title{Anisotropic Universe Models in $f(T)$ Gravity}
\author{M. E. Rodrigues $^{(a)}$\footnote{E-mail
address: esialg@gmail.com}, M. J. S. Houndjo $^{(b)(c)}$\footnote{E-mail address:
sthoundjo@yahoo.fr}, D. S\'aez-G\'omez$^{(d)}$\footnote{E-mail: diego.saez@ehu.es} and F. Rahaman $^{(e)}$\footnote{E-mail address: 
rahaman@iucaa.ernet.in} }
\medskip
\affiliation{$^{(a)}$Universidade Federal do Esp\'{\i}rito Santo \\
Centro de Ci\^{e}ncias
Exatas - Departamento de F\'{\i}sica\\
Av. Fernando Ferrari s/n - Campus de Goiabeiras - CEP29075-910 -
Vit\'{o}ria/ES, Brazil}
\affiliation{$^{(b)}$ Departamento de Ci\^{e}ncias Naturais - CEUNES\\
Universidade Federal do Esp\'irito Santo - 
CEP 29933-415 - S\~ao Mateus - ES, Brazil}
\affiliation{$^{(c)}$Institut de Math\'{e}matiques et de Sciences Physiques (IMSP) - 01 BP 613 Porto-Novo, B\'{e}nin}
\affiliation{$^{(d)}$Fisika Teorikoaren eta Zientziaren Historia Saila, Zientzia eta Teknologia Fakultatea,\\
Euskal Herriko Unibertsitatea, 644 Posta Kutxatila, 48080 Bilbao, Spain, EU}
\affiliation{$^{(e)}$Department of Mathematics, Jadavpur University, Kolkata - 700032, India}

\begin{abstract}
We investigate the cosmological reconstruction in anisotropic  universe for both  homogeneous and inhomogeneous content of the universe. Special attention is attached to three interesting cases: Bianchi type-I, Bianchi type-III and Kantowski-Sachs models. The de Sitter, power-law and general exponential solutions are assumed for the scale factor in each spatial direction and the corresponding cosmological models are reconstructed. Moreover, for the general exponential solutions, from which the de Sitter and power-law solutions may be obtained, we obtain models which reproduce the early universe, assumed as the inflation, and the late time accelerated expanding universe. The models obtained for the late time universe are consistent with a known result in literature where a power-law type correction in $T$ is added to a power-law type of $f(T)$ for guaranteeing the avoidance of the Big Rip and the Big Freeze. 
\end{abstract}
\maketitle 

\section{Introduction}


The probably presence of an unknown form of energy in the universe, called dark energy, and confirmed by a large number of observations, starting by the data of Supernovae IA in 1998 \cite{riess}, have leaded to explore the possible theoretical origin for this fluid. Since it is the direct responsible of the present accelerating expansion of the universe, a negative pressure is required which leads to a negative equation of state parameter. The most popular candidate, the cosmological constant, which posses a constant equation of state (EoS), $p_{\Lambda}=-\rho_{\Lambda}$, can explain quite well the cosmological evolution. However,  the open possibility that the EoS is not completely constant but evolutes dynamically (even crossing the phantom barrier more), and the quite large difference between the observed dark energy density and the vacuum energy density predicted by quantum field theories, have leaded to explore other possibilities, as the existence of scalar fields, vector fields or modifications of General Relativity (GR), among others (for a review on dark energy candidates, see \cite{copeland}). 
\par
In the context of modified gravities, a wide range of possibilities have been explored, being $f(R)$ gravity probably the most popular one due to its simplicity since it generalizes the Hilbert-Einstein action to a more complex function of the Ricci scalar (for a review on $f(R)$ gravity, see \cite{reviews,Bamba:2012cp}). Nevertheless, other kind of theories have been suggested, where other curvature invariants are included as the Gauss-Bonnet gravity. In this paper, we study the so-called $f(T)$ gravity, which in analogy to f(R) gravity, consists in a generalization of the action of Teleparallel gravity, a theory that assumes Weitzenbock connection instead of the Levi-Civita connection, which yields to a null curvature but a non-vanishing torsion (for a review see \cite{pereira1}). In this gravitational theory, the main field is represented by the so-called tetrads instead of the metric as in GR. This kind of theories has become very popular recently as can also explain the accelerated expansion of the universe with no need of dark energy, and even the inflationary epoch (see \cite{ferraro1}-\cite{ratbay}). Then, a wide number of aspects have been studied in the context of $f(T)$ gravity, as  its local Lorentz invariance \cite{barrow}, static solutions \cite{daouda1}, non-diagonal tetrads \cite{daouda3}, or the presence of wormholes \cite{boehmer}, as well as other aspects \cite{x,cemsinan}. Also a large effort has been done to study cosmological solutions for this class of theories, as well as possible cosmological predictions (see Refs.~\cite{houndjo}-\cite{li}).
\par
At the present work, we are interested to study some particular cosmological solutions in $f(T)$ gravity, where the appropriate action is reconstructed for each case. Specifically, the Bianchi type-I, Kantowski-Sachs (KS) and Bianchi type-III models are considered, and particularly some important solutions, such as power law and de Sitter (dS) expansion, or more complex ones as exponential functions for the scale factor in each direction of the space. Since power law and dS solutions can provide a good description for some specific phases of the universe evolution, their reconstruction in $f(T)$ gravity becomes a crucial point in order to consider this class of theories as serious candidates for explaining the whole cosmological history. In addition, here we assume more general cosmological metrics than Friedmann-Lema\^itre-Robertson-Walker (FLRW) metrics,  in particular anisotropic universes described by the Bianchi type-I, Kantowski-Sachs (KS) and Bianchi type-III metrics, in order to provide the most general description of the cosmological evolution in the context of $f(T)$ gravity.  Moreover, exponential solutions are also considered, this kind of expansions has become very popular recently as they may conduct the universe to a non-singular state, where some bounded systems may be broken. Such state suggested in Ref.~\cite{LittleRip}, and called {\it Little Rip},  has already been studied in $f(R)$ gravity (see Ref.~\cite{Nojiri:2011kd}), as well as in $f(T)$ theories \cite{bambamyr}. Even more,  the possible occurrence of a   {\it Little Rip} has been also explored in the context of the so-called viable modified gravities (see Ref.~\cite{SaezGomez:2012ek}). Note that anisotropic
cosmological metrics have been already studied in the context of GR with the presence of isotropic and anisotropic fluids, as well as the stability of the solutions \cite{barrow3,stability}.
\par
Furthermore, the use of an auxiliary scalar field, in analogy to the equivalence of Brans-Dicke theories for $f(R)$ gravity (see for instance Ref.~\cite{STFR}), is also implemented, from which may result a useful tool to reconstruct the appropriate action as well as for studying the properties of $f(T)$ gravity. 
\par
The main motivations of assuming  the assumption of a model with anisotropic geometry are based on several physical aspects as: the famous problem of the CMB quadrupole can be solved by considering a universe with planar symmetry \cite{campanelli1} where eccentricity in decoupling, generated by a uniform cosmic magnetic field whose current strength,  $ B(t_0) \sim 10^{-9}$ Gauss, should be close to $e_{dec}\sim 10^{-2}$; the Bianchi type models in Loop Quantum Cosmology \cite{lqc}, $^4He$ abundance \cite{campanelli2}, cosmic parallax \cite{campanelli3,fontanini}, small anisotropic pressures \cite{barrow2}, cosmological solutions of the low energy string effective action \cite{massimo}, anisotropic inflationary universe \cite{inflation} and some other \cite{x-1}. In the $f(R)$ theory, we already have some good results \cite{can,sharif2}, therefore, we propose to establish the equations here and get the first results in $f(T)$ gravity, for the Bianchi type-I, type-III and KS models.
\par
Then, the paper is organized as follows: in section \ref{sec2}, the basic concepts of $f(T)$ gravity are introduced. In section \ref{BK} , the equations for general  Bianchi type-I, type-III and Kantowski-Sachs (KS) models are deduced in a particular coordinate system and diagonal tetrads. Section \ref{sec2a} deals with the reconstruction of the $f(T)$ action for some relevant solutions, and where several techniques are considered, including a kind of scalar-tensor theory for torsion gravity. Finally, section \ref{conclusions} is devoted to the conclusions and discussions on the results found in the paper.
\section{\large Preliminary definitions and equations of motion}\label{sec2}


As previously mentioned, the $f(T)$ theory of gravity is defined in the Weitzenbock's space time in which the line element is described by 
\begin{equation}\label{el}
dS^{2}=g_{\mu\nu}dx^{\mu}dx^{\nu}\; ,
\end{equation} 
where $g_{\mu\nu}$ are the components of the metric which is symmetric and possesses $10$ degrees of freedom. One can describe the theory in the spacetime or in the tangent space, which allows us to rewrite the line element (\ref{el}) as follows 
\begin{eqnarray}
dS^{2} &=&g_{\mu\nu}dx^{\mu}dx^{\nu}=\eta_{ij}\theta^{i}\theta^{j}\label{1}\; ,\\
dx^{\mu}& =&e_{i}^{\;\;\mu}\theta^{i}\; , \; \theta^{i}=e^{i}_{\;\;\mu}dx^{\mu}\label{2}\; ,
\end{eqnarray} 
where $\eta_{ij}=diag[1,-1,-1,-1]$ and $e_{i}^{\;\;\mu}e^{i}_{\;\;\nu}=\delta^{\mu}_{\nu}$ or  $e_{i}^{\;\;\mu}e^{j}_{\;\;\mu}=\delta^{j}_{i}$. The square root of the metric determinant is given by  $\sqrt{-g}=\det{\left[e^{i}_{\;\;\mu}\right]}=e$ and the matrix $e^{a}_{\;\;\mu}$ are called tetrads and represent the dynamic fields of the theory.
\par
By using theses fields, one can define the Weitzenbock's connection as 
\begin{eqnarray}
\Gamma^{\alpha}_{\mu\nu}=e_{i}^{\;\;\alpha}\partial_{\nu}e^{i}_{\;\;\mu}=-e^{i}_{\;\;\mu}\partial_{\nu}e_{i}^{\;\;\alpha}\label{co}\; .
\end{eqnarray}
The main geometrical objects of the spacetime are constructed from this connection. The components of the tensor torsion are defined by the antisymmetric part of this connection
\begin{eqnarray}
T^{\alpha}_{\;\;\mu\nu}&=&\Gamma^{\alpha}_{\nu\mu}-\Gamma^{\alpha}_{\mu\nu}=e_{i}^{\;\;\alpha}\left(\partial_{\mu} e^{i}_{\;\;\nu}-\partial_{\nu} e^{i}_{\;\;\mu}\right)\label{tor}\;.
\end{eqnarray}
The components of the contorsion are defined as 
\begin{eqnarray}
K^{\mu\nu}_{\;\;\;\;\alpha}&=&-\frac{1}{2}\left(T^{\mu\nu}_{\;\;\;\;\alpha}-T^{\nu\mu}_{\;\;\;\;\alpha}-T_{\alpha}^{\;\;\mu\nu}\right)\label{contor}\; .
\end{eqnarray}
In order to make more clear the definition of the scalar equivalent to the curvature scalar of RG, we first define a new tensor $S_{\alpha}^{\;\;\mu\nu}$, constructed from the components of the tensors torsion and contorsion as
\begin{eqnarray}
S_{\alpha}^{\;\;\mu\nu}&=&\frac{1}{2}\left( K_{\;\;\;\;\alpha}^{\mu\nu}+\delta^{\mu}_{\alpha}T^{\beta\nu}_{\;\;\;\;\beta}-\delta^{\nu}_{\alpha}T^{\beta\mu}_{\;\;\;\;\beta}\right)\label{s}\;.
\end{eqnarray}
We can now define the torsion scalar by the following contraction
\begin{eqnarray}
T=T^{\alpha}_{\;\;\mu\nu}S^{\;\;\mu\nu}_{\alpha}\label{te}\; .
\end{eqnarray}

The action of the theory is defined by generalizing  the Teleparallel theory, as 
\begin{eqnarray}\label{action}
S=\int e\left[f(T)+\mathcal{L}_{Matter}\right]d^4x\;,
\end{eqnarray}
where $f(T)$ is an algebraic function of the torsion scalar $T$. Making the functional variation of the action  (\ref{action}) with respect to the tetrads, we get the following equations of motion \cite{barrow,daouda1,daouda2}
\begin{eqnarray}
S^{\;\;\nu\rho}_{\mu}\partial_{\rho}Tf_{TT}+\left[e^{-1}e^{i}_{\mu}\partial_{\rho}\left(ee^{\;\;\alpha}_{i}S^{\;\;\nu\rho}_{\alpha}\right)+T^{\alpha}_{\;\;\lambda\mu}S^{\;\;\nu\lambda}_{\alpha}\right]f_{T}+\frac{1}{4}\delta^{\nu}_{\mu}f=4\pi\mathcal{T}^{\nu}_{\mu}\label{em}\; ,
\end{eqnarray}
where $\mathcal{T}^{\nu}_{\mu}$ is the energy momentum tensor, $f_{T}=d f(T)/d T$ and $f_{TT}=d^{2} f(T)/dT^{2}$. By setting $f(T)=a_1T+a_0$,  the equations of motion (\ref{em}) are the same as that of the Teleparallel theory with a cosmological constant, and this is dynamically equivalent to the GR.  These equations clearly depend on the choice made for the set of tetrads \cite{cemsinan}. 
\par
The contribution of the interaction with the matter fields is given by the energy momentum tensor which, is this case, is defined as 
\begin{eqnarray}
\mathcal{T}^{\,\nu}_{\mu}=  diag\left(1,-\omega_x,-\omega_y,-\omega_z\right)\rho                         \label{tem}\; ,
\end{eqnarray}
where the $\omega_i$ ($i=x,y,z$) are the parameters of equations of state related to the pressures $p_x$, $p_y$ and $p_z$.

\section{Field equations for Bianchi type-I, type-III and Kantowski-Sachs models} \label{BK}


Let us first establish the equations of motion of a set of diagonal tetrads using the Cartesian coordinate metric, for  describing models of Bianchi type-I, type-III and Kantowski-Sachs (KS). We propose to start with the Bianchi type-III case, from which Bianchi type-I and KS can be recovered. For the Bianchi type-III case, the metric reads 
\begin{equation}
dS^2=dt^2-A^2(t)dx^2-e^{-2\alpha x}B^2(t)dy^2-C^2(t)dz^2\,\,\,,\label{metrictype3}
\end{equation}
where $\alpha$ is a constant parameter. Note that the Bianchi type-I is recovered by setting $\alpha=0$ from the Bianchi type-III, while KS is recovered when one takes $\alpha=0$ and $B(t)=C(t)$. Let us choose the following set of diagonal tetrads related to the metric (\ref{metrictype3})
\begin{eqnarray}
\left[e^{a}_{\;\;\mu}\right]=diag\left[1,A,e^{-\alpha x}B,C\right]\;. \label{matrixtype3}
\end{eqnarray}
The determinant of the matrix (\ref{matrixtype3}) is $e=e^{-\alpha x}ABC$. The components of the tensor torsion (\ref{tor}) for the tetrads (\ref{matrixtype3}) are given by
\begin{eqnarray}
T^{1}_{\;\;01}=\frac{\dot{A}}{A}\,,\,T^{2}_{\;\;02}=\frac{\dot{B}}{B}\,,\, T^{2}_{\;\;21}=\alpha\,,\,T^{3}_{\;\;03}=\frac{\dot{C}}{C}\;,\label{torsiontype3}
\end{eqnarray}
and the components of the corresponding tensor contorsion are 
\begin{eqnarray}
K^{01}_{\;\;\;\;1}=\frac{\dot{A}}{A}\,,\,K^{02}_{\;\;\;\;2}=\frac{\dot{B}}{B}\,,\,K^{12}_{\;\;\;\;2}=\frac{\alpha}{A^2}\,,\,K^{03}_{\;\;\;\;3}=\frac{\dot{C}}{C}\;.\label{contorsiontype3}
\end{eqnarray} 
The components of the tensor $S_{\alpha}^{\;\;\mu\nu}$, in (\ref{s}),  are given by
\begin{eqnarray}
S_{0}^{\;\;01}=S_{3}^{\;\;31}=\frac{\alpha}{2A^2}\,,\,S_{1}^{\;\;10}=\frac{1}{2}\left(\frac{\dot{B}}{B}+\frac{\dot{C}}{C}\right)\,,\,S_{2}^{\;\;20}=\frac{1}{2}\left(\frac{\dot{A}}{A}+\frac{\dot{C}}{C}\right)\,,\,S_{3}^{\;\;30}=\frac{1}{2}\left(\frac{\dot{A}}{A}+\frac{\dot{B}}{B}\right)\;.\label{tensortype3}
\end{eqnarray}
By using the components (\ref{torsiontype3}) and (\ref{tensortype3}),  the torsion scalar (\ref{te}) is given by
\begin{eqnarray}
T=-2\left(\frac{\dot{A}\dot{B}}{AB}+\frac{\dot{A}\dot{C}}{AC}+\frac{\dot{B}\dot{C}}{BC}\right)\; \label{torsionScalar1}.
\end{eqnarray}
The equations of motion are given by 
\begin{eqnarray}
16\pi \rho &=& f+4f_T\Big[\frac{\dot{A}\dot{B}}{AB}+\frac{\dot{A}\dot{C}}{AC}+\frac{\dot{B}\dot{C}}{BC}-\frac{\alpha^2}{2A^2}\Big]\,\,,\label{densitytype3}\\
-16\pi p_x &=& f+2f_T\left[\frac{\ddot{B}}{B}+\frac{\ddot{C}}{C}+\frac{\dot{A}\dot{B}}{AB}+\frac{\dot{A}\dot{C}}{AC}+2\frac{\dot{B}\dot{C}}{BC}\right]\nonumber\\&+&2\left(\frac{\dot{B}}{B}+\frac{\dot{C}}{C}\right)\dot{T}f_{TT}\;,\label{radialpressuretype3}\\
-16\pi p_y&=&f+2f_T\left[\frac{\ddot{A}}{A}+\frac{\ddot{C}}{C}+\frac{\dot{A}\dot{B}}{AB}+2\frac{\dot{A}\dot{C}}{AC}+\frac{\dot{B}\dot{C}}{BC}\right]\nonumber\\&+&2\left(\frac{\dot{A}}{A}+\frac{\dot{C}}{C}\right)\dot{T}f_{TT}\;,\label{tangentialpressure1type3}\\
-16\pi p_z=f&+&2f_T\left[\frac{\ddot{A}}{A}+\frac{\ddot{B}}{B}+2\frac{\dot{A}\dot{B}}{AB}+\frac{\dot{A}\dot{C}}{AC}+\frac{\dot{B}\dot{C}}{BC}-\frac{\alpha^2}{A^2}\right]\nonumber\\&+&2\left(\frac{\dot{A}}{A}+\frac{\dot{B}}{B}\right)\dot{T}f_{TT}\;,\label{tangentialpressure2type3}\\
\frac{\alpha}{2A^2}\left[\left(\frac{\dot{A}}{A}-\frac{\dot{B}}{B}\right)f_T-\dot{T}f_{TT}\right]&=&0 \,\,\,, \label{constraint1} \\
\alpha\left(\frac{\dot{A}}{A}-\frac{\dot{B}}{B}\right)f_T&=&0\;.\label{constraint2}
\end{eqnarray}
In the particular case where $f(T)=T-2\Lambda$, the equations (\ref{densitytype3})-(\ref{constraint2}) are identical to that of the GR \cite{akarsu}. The equation of constraint (\ref{constraint2}) appears in both the GR as in $f(R)$ gravity \cite{sharif2}. But here we have a second equation of constraint (\ref{constraint1}), which appears as a generalization of the previous one, because here we have a contribution of a term of second derivative of the function $f(T)$ with respect to $T$.
\par

By setting $\alpha=0$, the Bianchi type-I case is recovered and the equations of motions read
\begin{eqnarray}
16\pi \rho &=& f+4f_T\Big[\frac{\dot{A}\dot{B}}{AB}+\frac{\dot{A}\dot{C}}{AC}+\frac{\dot{B}\dot{C}}{BC}\Big]\,\,\,,\label{densitytype1}\\
-16\pi p_x &=& f+2f_T\left[\frac{\ddot{B}}{B}+\frac{\ddot{C}}{C}+\frac{\dot{A}\dot{B}}{AB}+\frac{\dot{A}\dot{C}}{AC}+2\frac{\dot{B}\dot{C}}{BC}\right]+2\left(\frac{\dot{B}}{B}+\frac{\dot{C}}{C}\right)\dot{T}f_{TT}\;,\label{radialpressuretype1}\\
-16\pi p_y&=&f+2f_T\left[\frac{\ddot{A}}{A}+\frac{\ddot{C}}{C}+\frac{\dot{A}\dot{B}}{AB}+2\frac{\dot{A}\dot{C}}{AC}+\frac{\dot{B}\dot{C}}{BC}\right]+2\left(\frac{\dot{A}}{A}+\frac{\dot{C}}{C}\right)\dot{T}f_{TT}\;,\label{tangentialpressure1type1}\\
-16\pi p_z&=&f+2f_T\left[\frac{\ddot{A}}{A}+\frac{\ddot{B}}{B}+2\frac{\dot{A}\dot{B}}{AB}+\frac{\dot{A}\dot{C}}{AC}+\frac{\dot{B}\dot{C}}{BC}\right]+2\left(\frac{\dot{A}}{A}+\frac{\dot{B}}{B}\right)\dot{T}f_{TT}\;\label{tangentialpressure2type1}\;.
\end{eqnarray}
The equations of motion corresponding to KS model are obtained by setting $\alpha=0$ and $B=C$, yielding
\begin{eqnarray}
16\pi\rho&=&f+4f_T\left[\left(\frac{\dot{B}}{B}\right)^2+2\frac{\dot{A}\dot{B}}{AB}\right]\;,\label{densityks}\\
-16\pi p_x&=&f+4f_T\left[\frac{\ddot{B}}{B}+\left(\frac{\dot{B}}{B}\right)^2+\frac{\dot{A}\dot{B}}{AB}\right]+4\frac{\dot{B}}{B}\dot{T}f_{TT}\;,\label{radialpressureks}
\end{eqnarray}
\begin{eqnarray}
-16\pi p_y&=&f+2f_T\left[\frac{\ddot{A}}{A}+\frac{\ddot{B}}{B}+\left(\frac{\dot{B}}{B}\right)^2+3\frac{\dot{A}\dot{B}}{AB}\right]+2\left(\frac{\dot{A}}{A}+\frac{\dot{B}}{B}\right)\dot{T}f_{TT}\;,\label{tangentialpressureks}\\
p_y&=&p_z\nonumber\,\,\,.
\end{eqnarray}
In the next section we will perform the reconstruction scheme of the action of the system for some particular cases.

\section{ Reconstructing $f(T)$ gravity in inhomogeneous universes}\label{sec2a}


Let us now consider the reconstruction of the $f(T)$ action for some particular solutions of the class of metrics explored in the previous section. Specifically, we consider solutions of the type of de Sitter, power law evolutions and exponential solutions. Note that de Sitter and power law solutions have been widely explored in other contexts of modified gravity, as $f(R)$ and Gauss-Bonnet gravities (see Ref.~\cite{Nojiri:2009kx}), since they can provide a well description of  the cosmological evolution along its particular phases.\\
Let's start by considering for simplicity Bianchi type-I and Kantowski-Sachs $(\alpha=0)$ metrics. Then, the conservation equation for the energy momentum tensor (\ref{tem}) can be easily obtained,
\be
\dot{\rho}+\left(H_x+H_y+H_z\right)\rho+H_x p_x+H_y p_y +H_z p_z=0\ ,
\label{D1}
\ee
where we have defined $H_x=\frac{\dot{A}}{A}\ \ H_y=\frac{\dot{B}}{B}\ \ H_z=\frac{\dot{C}}{C}$. We can now analyse de Sitter, power law solutions and exponential expansion in  Bianchi type-I metric by one side, and Kantowski-Sachs metric by the other, where $B=C$ that implies $p_y=p_z$.

\subsection{De Sitter solutions}

De Sitter solutions are well known in the context of cosmology since the current epoch, where the universe expansion is being accelerated, can be described approximately with a de Sitter solution. This kind of solutions consists on an exponential expansion of the scale factor, which yields a constant Hubble parameter. In the case of Bianchi type-I and Kantowski-Sachs metrics $(\alpha=0)$ in (\ref{metrictype3}), we may assume an exponential expansion for each spatial direction,
\be
A=A_0 \e^{a t}\ \ B=B_0 \e^{b t}\ \ C=C_0 \e^{c t}\ ,
\label{D2}  
\ee
and the rates of the expansion for each direction can be defined as,
\be
H_x=\frac{\dot{A}}{A}=H_{x0}\ \ H_{y}=\frac{\dot{B}}{B}=H_{y0}\ \ H_c=\frac{\dot{C}}{C}=H_{z0}\ ,
\label{D3}
\ee
where $\{H_{x0}=a,H_{y0}=b,H_{z0}=c\}$ are constants. The torsion scalar defined in (\ref{torsionScalar1}) is given by,
\be
T_0=-2\left(H_{x0}H_{y0}+H_{x0}H_{z0}+H_{y0}H_{z0}\right)\ .
\label{D4}
\ee
Then, by assuming $p_x=p_y=p_z=p$ and an equation of state $p=w\rho$, the conservation equation (\ref{D1}) can be easily solved for the ansatz  (\ref{D2}),
\be
\rho=\rho_0\e^{-(H_{x0}+H_{y0}+H_{z0})(1+w)t}\ .
\label{D5}
\ee
Hence, the field equations (\ref{densitytype1})-(\ref{tangentialpressure2type1}) become,
\bea
16\pi \rho_0\e^{-(H_{x0}+H_{y0}+H_{z0})(1+w)t}=f(T_0)+4\left[H_{x0}H_{y0}+H_{z0}(H_{x0}+H_{y0})\right]f_T(T_0)\ , \label{D6} \\
-16\pi w\rho_0\e^{-(H_{x0}+H_{y0}+H_{z0})(1+w)t}=f(T_0)+2(H_{y0}+H_{z0})(H_{x0}+H_{y0}+H_{z0})f_T(T_0)\ , \label{D7}\\
-16\pi w\rho_0\e^{-(H_{x0}+H_{y0}+H_{z0})(1+w)t}=f(T_0)+2(H_{x0}+H_{z0})(H_{x0}+H_{y0}+H_{z0})f_T(T_0)\ , \label{D8}\\
-16\pi w\rho_0\e^{-(H_{x0}+H_{y0}+H_{z0})(1+w)t}=f(T_0)+2(H_{x0}+H_{y0})(H_{x0}+H_{y0}+H_{z0})f_T(T_0)\ . \label{D9}
\eea
Note that the only possible solution in the presence of a perfect fluid is one with $w=-1$ as the r.h.s. of equations (\ref{D6})-(\ref{D9}) is independent of time, according to the expression of the scalar torsion for a pure de Sitter solution (\ref{D4}), unless $H_{x0}+H_{y0}+H_{z0}=0$, which would imply a decelerating expansion in a particular direction, being $H_{i0}<0$. Moreover, for a particular $f(T)$ action, the system of equations (\ref{D4})-(\ref{D9}) reduces to an algebraic system of equations for the variables $\{H_{x0},H_{y0},H_{z0}\}$. Since  the system of equations (\ref{D4})-(\ref{D9}) are composed by four equations, while there are only three variables,
the above 4-equations system has to be reduced. However, even in the case of  Kantowski-Sachs metric, where $B(t)=C(t)\rightarrow H_{y0}=H_{z0}$, the system (\ref{D4})-(\ref{D9}) still posses three independent equations with two variables. Hence, the only possible solution imposes,
\be
 A(t)=B(t)=C(t)\rightarrow H_{x0}=H_{y0}=H_{z0}=H_0\ ,
 \label{D10}
\ee
And the metric (\ref{metrictype3}) reduces to the well known Friedmann-Lema\^itre-Robertson-Walker metric with an exponential expansion, $A(t)=A_0\e^{H_0\ t}$. Hence, the only solution for a pure de Sitter expansion in  Bianchi type-I and Kantowski-Sachs metrics gives a FLRW universe\footnote{Recall that we have assumed here that the pressures are equal, $p_x=p_y=p_z$.}, and the system of equations (\ref{D4})-(\ref{D9}) reduces now to a unique independent equation,
 \be
 16 \pi \rho_0=f(T_0)+12H_{0}^2f_T(T_0)\ .
 \label{D11}
 \ee
 Then, the roots of the algebraic equation (\ref{D11}) give the de Sitter points of a particular $f(T)$ action. In order to illustrate such possibility, let us consider the action,
 \be
 f(T)=\left(-T\right)^n\ ,
 \label{D12}
 \ee
 where $n$ is a real constant. Then, the equation (\ref{D11}) is rewritten as,
 \be
 16\pi \rho_0=(1-2n)(6H_{0}^2)^n\ ,
 \label{D13}
 \ee
 whose solution is given by,
 \be
 H_{0}^2=\frac{1}{6}\left(\frac{16\pi\rho_0}{1-2n}\right)^{1/n}\ .
 \label{D14}
\ee
 Hence, the only physical solution ($\rho_0,H_0^2\geq0$) imposes $n\leq1/2$. Then, the de Sitter solution is a direct consequence of the energy density $\rho_0$, which can be interpreted as a cosmological constant according to the condition imposed above for its equation of state, $w=-1$. Nevertheless, in vacuum the equation (\ref{D13}) reduces to $0=(1-2n)(6H_{x0}^2)^n$, whose only solution is given by $n=1/2$, rising to $f(T)=\sqrt{-T}$ that posses an infinite number of de Sitter points. Moreover, we may consider in vacuum the action,
 \be
 f(T)=C_1 T+C_2 \left(-T\right)^n\ ,
 \label{D15}
 \ee
where $\{C_1,C_2\}$ are the coupling constants and $n$ is a real constant. The field equation (\ref{D11}) in vacuum yields,
 \be
0= C_1 6H_{0}^2+C_2(1-2n)(6H_{0}^2)^n\ .
\label{D16}
\ee
So the roots of this equation give the dS points allowed by the class of theories expressed in (\ref{D15}). Note that now, the exponential expansion is a direct consequence of the action instead of the contribution of a kind of cosmological constant as in the case shown above. For instance, $n=2$, it yields the solution,
\be
H_{0}=\sqrt{\frac{C_1}{18C_2}}\ . 
\label{D17}
\ee
While for higher powers of $n$, more de Sitter points can be obtained for the action (\ref{D15}). Note that in $f(R)$ theories, dS points constitutes the critical points of the dynamical system, which may be (un)stable, and could explain both the inflationary and dark energy epochs (see \cite{Cognola:2008zp}), which may be the case also in $f(T)$ gravity.

\subsection{Power law solutions}

Let us now explore a cosmological evolution described by a power law in each direction of the space expansion. In such case, the scale parameters for the Bianchi type-I and Kantowski-Sachs  metric (\ref{metrictype3}), where we set $(\alpha=0)$, can be expressed as,
\be
A(t)=A_0t^a\ , \ B(t)=B_0t^b\ , \ C(t)=C_0t^c\ ,
\label{D18}
\ee
where $\{a,b,c\}$ and $\{A_0,B_0,C_0\}$ are constants to be determined by the field equations, and initial conditions respectively. The expansion rates are given by,
\be
H_x=\frac{a}{t}\ , \ H_y=\frac{b}{t}\ , \ H_z=\frac{c}{t}\ .
\label{D19}
\ee
While the expression for the torsion scalar (\ref{torsionScalar1}) yields,
\be
T=-2\left(\frac{ab}{t^2}+\frac{ac}{t^2}+\frac{bc}{t^2}\right)\ .
\label{D20}
\ee
Then, introducing the above quantities in the field equations (\ref{densitytype1})-(\ref{tangentialpressure2type1}), we get the following system of differential equations in $f(T)$,
\bea
16\pi \rho(T)=f(T)-2Tf_T(T)\ , \label{D21} \\
-16\pi w\rho(T)=f(T)+\frac{(b+c)(1-a-b-c)}{bc+a(b+c)}Tf_T(T)+2\frac{(b+c)}{bc+a(b+c)}T^2f_{TT}(T)\ , \label{D22}\\
-16\pi w\rho(T)=f(T)+\frac{(a+c)(1-a-b-c)}{bc+a(b+c)}Tf_T(T)+2\frac{(a+c)}{bc+a(b+c)}T^2f_{TT}(T)\ ,\label{D23} \\
-16\pi w\rho(T)=f(T)+\frac{(a+b)(1-a-b-c)}{bc+a(b+c)}Tf_T(T)+2\frac{(a+b)}{bc+a(b+c)}T^2f_{TT}(T)\ , \label{D24}
\eea
where we have assumed for simplicity that $p_x=p_y=p_z=p$ and an EoS $p=w\rho$. Hence, the system (\ref{D21})-(\ref{D24}) is a set of differential equations in f(T) with the torsion scalar $T$ as the independent variable. \\ 

Firstly let us consider the vacuum case, or in other words, the homogeneous part of the first equation (\ref{D21}), which becomes $f(T)-2Tf_T(T)=0$ and whose solution is given by,
\be
f(T)=C_1 \sqrt{-T}\ ,
\label{D26}
\ee
where $C_1$ is an integration constant. In order to satisfy the rest of the equations (\ref{D22})-(\ref{D24}), the condition $a=b=c$ must be imposed, so that the Hubble parameters (\ref{D19}) reduce to the usual FLRW cosmology reproducing  power law solution. \\

In the presence of a kind of isotropic perfect fluid $p=w\rho$, we can first solve the continuity equation (\ref{D1}) in order to obtain $\rho=\rho(T)$, which yields,
\be
\rho=\rho_0 t^{-(a+b+c)(1+w)}=\rho_0 \left(-\frac{T}{2(ab+ac+bc)}\right)^{\frac{(a+b+c)(1+w)}{2}}\ .
\label{D25}
\ee
Hence, the solution for the set of equations (\ref{D21})-(\ref{D24}) is given by $f(T)=f_h(T)+f_p(T)$, where $f_h(T)$ is the solution of the homogeneous equation, which coincides with the vacuum solution (\ref{D26}), and $f_p(T)$ is the particular solution. Then, by using (\ref{D25}) in the equation (\ref{D21}), the particular solution can be easily found,
\be
f_{p}(T)=\chi\ T^{\frac{(1+w)(a+b+c)}{2}}\ , 
\label{D27} 
\ee
 where $\chi$ is a constant given by
\be
 \chi=\frac{2^{4-(1+w)(a+b+c)/2}\pi \rho_0}{\left[-1+(1+w)(a+b+c)\right] \left[-bc-a(b+c)\right]^{\frac{(1+w)(a+b+c)}{2}}}\ .
 \label{D28}
\ee
Note that the condition $(1+w)(a+b+c)=2\ n$ with $n$ being any natural number,  has to be imposed in order to avoid a complex gravitational action that would lacks any physical sense, recall  that $T<0$ for an expanding universe according to (\ref{D20}).   In order to satisfy the complete set of equations (\ref{D21})-(\ref{D24}), we introduce the solution (\ref{D27}) into the field equations (\ref{D22})-(\ref{D24}), and the following solutions for the parameters  $\{a,b,c\}$ are found,
\begin{enumerate}[ i.]
\item $\ c=\frac{1-w(a+b)}{w}\ $, where $w\neq 0$. This provides an anisotropic solution in $f(T)$ gravity with $A(t)$, $B(t)$ and $C(t)$ being different functions in (\ref{D18}), and recalling that the perfect fluid assumed is an isotropic fluid. This does not hold in GR or Teleparallel Theory (TT) but is possible here due to the presence of second derivatives of the function $f(T)$ with respect to the torsion scalar $T$ in (\ref{D21})-(\ref{D24}). Note that field equations may be rewritten as the usual equations in TT, 
\[
H_xH_y+H_xH_z+H_yH_z=16\pi(\rho +\rho_{f(T)})\ ,\; \;
-\dot{H}_y-H_{y}^2-\dot{H}_z-H_{z}^2-H_y\ H_z=8\pi (w\rho +p_{f(T)}^x)\ ,
\]
\be
-\dot{H}_x-H_{x}^2-\dot{H}_z-H_{z}^2-H_x\ H_z=8\pi (w\rho +p_{f(T)}^y)\ ,\; \;
-\dot{H}_x-H_{x}^2-\dot{H}_y-H_{y}^2-H_x\ H_y=8\pi (w\rho +p_{f(T)}^z)\ .
\ee
Here, the extra terms coming from $f(T)$ are defined as an energy density $\rho_{f(T)}$ and pressures $\{p_{f(T)}^x, \ p_{f(T)}^y, \ p_{f(T)}^z\}$, which are the origin of the anisotropic evolution. In this case, we have to fix $C_1=0$  in (\ref{D26}) in order to satisfy the whole system.
\item $a=b=c$. The cosmological evolution expressed by \ref{D18} reduces to a FLRW metric as in the homogeneous part of the equations, so that $C_1\neq0$. 
\item $c=-ab/(a+b)$. In spite of this satisfies equations (\ref{D21})-(\ref{D24}) once (\ref{D27}) is substituted in the equations, it gives $T=0$, and the r.h.s of equations (\ref{D21})-(\ref{D24}) become null while the l.h.s. are not, since $\rho=\rho(t)$ as given in (\ref{D25}), so this is not a real solution.

\end{enumerate}�
\par
Therefore,  we have obtained a complete set of power law solutions for Bianchi type-I  universe and Kantowski-Sachs metrics in the context of $f(T)$ gravity. Nevertheless, the action is clearly dependent on the EoS parameter $w$. Note also that in vacuum, the only possible solution reduces to a FLRW metric.

\subsection{General exponential solutions}

In this subsection we consider a more general exponential expansion for each spatial direction by 
\begin{eqnarray}
A=A_0e^{g_x(t)}\,\,\,, \quad B=B_0e^{g_y(t)}\,\,\,,\quad C=C_0e^{g_z(t)}\,\,\,,
\end{eqnarray}
where the function $g_i(t)$ is assumed as 
\begin{eqnarray}
g_i(t)=h_i(t)\ln{\left(t\right)}\,\,,\quad i=x,y,z\;\;,
\end{eqnarray}
 and $A_0$, $B_0$ and $C_0$ are positive constants. Note that the previous cases, the de Sitter solutions and power law solutions can be recovered from this one by setting $h_i(t)=a_it/(\ln{(t)})$ and $h_i(t)=a_i$, respectively, where $\{a_i\}=\{a,b,c\}$. In what follows, we will use an adiabatic approximation for the expansion in each spatial direction and neglect the derivatives of $h_i(t)$, i.e., setting $(\dot{h}_i\sim \ddot{h}_i\sim 0)$. The expansion rates in this case are given by
\begin{eqnarray}
H_x=\frac{h_x(t)}{t}\,\,,\quad H_y=\frac{h_y(t)}{t}\,\,,\quad H_z=\frac{h_z(t)}{t}\,\,.
\label{steph62}
\end{eqnarray}
Thus, the torsion scalar (\ref{torsionScalar1}) becomes
\begin{eqnarray}
T=-2\left[\frac{h_x(t)h_y(t)}{t^2}+\frac{h_x(t)h_z(t)}{t^2}+\frac{h_y(t)h_z(t)}{t^2}\right]\,\,.\label{steph63}
\end{eqnarray} 
The acceleration in each direction is given by
\begin{eqnarray}
\ddot{A}=\frac{h_x(h_x-1)}{t^2}A\,\,,\quad \ddot{B}=\frac{h_y(h_y-1)}{t^2}B\,\,,\quad \ddot{C}=\frac{h_z(h_z-1)}{t^2}C\,\,.\label{steph64}
\end{eqnarray}
Since $A$, $B$ and $C$ are positives, the acceleration is guaranteed in each direction when $h_i>1$, while for $0<h_i<1$, the universe  is in deceleration.\par
The simplest example of $h_i(t)$ is
\begin{eqnarray}
h_i(t)=\frac{h_{i\,in}+h_{i\,out}\,qt^2}{1+qt^2}\,\,\,,\label{steph65}
\end{eqnarray}
where $h_{i\,in}$, $h_{i,out}$ and $q$ are positive constants, and the $q$ is assumed to be enough small in order to make $h_i(t)$ varying slowly. Thus, the torsion scalar is always negative. From (\ref{steph65}), we see that at early time $t=0$, $h_i\rightarrow h_{i\,in}$ and for late universe, $h_i\rightarrow h_{i\,out}$. By using 
(\ref{steph65}) and (\ref{steph63}), one gets the following equation
\begin{eqnarray}
q^2Tt^6+\left(2qT+2q^2X\right)t^4+\left(T+2qY\right)t^2+2Z=0\,\,\,,\\
X=h_{x\,out}h_{y\,out}+h_{x\,out}h_{z\,out}+h_{y\,out}h_{z\,out}\,\,,\quad
Y=h_{x\,in}h_{y\,out}+h_{x\,out}h_{y\,in}+h_{x\,in}h_{z\,out}+\nonumber\\
h_{x\,out}h_{z\,in}+h_{y\,in}h_{z\,out}+h_{y\,out}h_{z\,in}\,\,,\quad 
Z=h_{x\,in}h_{y\,in}+h_{x\,in}h_{z\,in}+h_{y\,in}h_{z\,in}\,\,,\nonumber
\end{eqnarray}
whose solutions read
\begin{eqnarray}
t^2&=&\left\{\Psi_0(T),\;\;\Psi_{\pm}(T)\right\}\,,\label{steph67}\\
\Psi_0(T)&\equiv& \alpha^{1/3}+\beta_0\alpha^{-1/3}+\beta_1\,\;\;\Psi_{\pm}(T)\equiv e^{\mp 2i\pi/3}\alpha^{1/3}+e^{\pm 2i\pi/3}\alpha^{-1/3}+\beta_1\,\,,\;\;\,\nonumber\\ \alpha&=&\frac{\sqrt{\alpha_1}}{q^2T^2}-\alpha_2\,\,,\;
\beta_0=\frac{qT(8X-6Y)+4q^2X^2+T^2}{9q^2T^2}\,\,,\quad \beta_1=-\frac{2qX+2T}{3qT}\,\,,\nonumber\\
\alpha_1&=&\frac{1}{27qT}\Big[27qT^2Z^2+[(-36q^2TX-36qT^2)Y+16q^3X^3+48q^2TX^2+30qT^2X-2T^3]Z\nonumber
\\&+&8q^2TY^3+(-4q^3X^2-8q^2TX+8qT^2)Y^2+(-4q^2TX^2-8qT^2X+2T^3)Y-qT^2X^2-2T^3X\Big]\,\,,\nonumber\\
\alpha_2&=&\frac{qT^2(27Z-18Y+15X)+q^2T(24X^2-18XY)+8q^3X^3-T^3}{27q^3T^3}\,\,.\nonumber
\end{eqnarray}
We see from (\ref{steph67}) that there are one real positive solution, $\Psi_0(T)$, and two complex solutions $\Psi_{\pm}(T)$. By using (\ref{steph62})-(\ref{steph64}) the system of equations of motion (\ref{densitytype1})-(\ref{tangentialpressure2type1}) becomes
\begin{eqnarray}
16\pi\rho&=&f-2Tf_T\,\,\,,\label{steph68}\\
-16\pi p_x&=&f+2f_T\left[\frac{(h_y+h_z)^2-(h_y+h_z)-h_yh_z}{\Psi_0(T)}-\frac{T}{2}\right]+4\left(\frac{h_y+h_z}{\Psi_0(T)}\right){T}f_{TT}\,\,,\label{steph69}\\
-16\pi p_y&=&f+2f_T\left[\frac{(h_x+h_z)^2-(h_x+h_z)-h_xh_z}{\Psi_0(T)}-\frac{T}{2}\right]+4\left(\frac{h_x+h_z}{\Psi_0(T)}\right){T}f_{TT}\,\,,\label{steph70}\\
-16\pi p_z&=&f+2f_T\left[\frac{(h_x+h_y)^2-(h_x+h_y)-h_xh_y}{\Psi_0(T)}-\frac{T}{2}\right]+4\left(\frac{h_x+h_y}{\Psi_0(T)}\right){T}f_{TT}\,\,,\label{steph71}\\
h_i&=&\frac{h_{i in}+h_{i out}q\Psi_0(T)}{1+q\Psi_0(T)}\,\,\,.\label{steph72}
\end{eqnarray} 
\subsubsection{A special case}\label{numero1}

Here, we assume that the expansion rates are equal in the three spatial directions ($h_x=h_y=h_z$), and the system (\ref{steph68})-(\ref{steph71}) reduces to
\begin{eqnarray}
16\pi\rho&=&f-2Tf_{T}\,\,,\label{steph73}\\
-16\pi p_x&=&f-2f_T\left(T+\sqrt{\frac{-2T}{3\Psi_0(T)}}\right)+4T\sqrt{\frac{-2T}{3\Psi_0(T)}}f_{TT}\,\,,\label{steph74}\\
p_x&=&p_y=p_z\,\,\,,\nonumber
\end{eqnarray}
which means that the assumption of having the same rate in the three spatial direction leads to an isotropic matter content. By combining (\ref{steph73}) and (\ref{steph74}), one gets 
\begin{eqnarray}
4T\sqrt{\frac{-2T}{3\Psi_0(T)}}f_{TT}-2f_T\left[(1+\omega)T+\sqrt{\frac{-2T}{3\Psi_0(T)}}\,\right]+(1+\omega)f=0\,\,\,,\label{steph75}
\end{eqnarray}
where we used the barotropic equation $p_x=\omega\rho$. Let us consider an asymptotic analysis, looking for the early universe (small time) and the late time one (large time), for which the function $h_i(t)$ yields $h_{i\,in}$ and $h_{i\,out}$, respectively. Thus, for the early universe Eq. (\ref{steph75}) reduces to 
\begin{eqnarray}
4T^2f_{TT}+6Tf_{T}\left[3h_{i\,in}(\omega+1)-1\right]-3h_{i\,in}(\omega+1)f=0\,\,\,,\\
h_{i\,in}=h_{x\,in}=h_{y\,in}=h_{z\,in}\,\,\,,\nonumber
\label{steph76}
\end{eqnarray}
whose general solution reads 
\begin{eqnarray}
f(T)=C_3T^{\gamma_+}+C_4T^{\gamma_-}\,\,,\quad \gamma_{\pm}=\frac{5-9h_{x\,in}(1+\omega)\pm\sqrt{25-78h_{x\,in}(1+\omega)+81h^2_{x\,in}(1+\omega)^2}}{4}\,\,,\label{steph77}
\end{eqnarray}
where $C_3$ and $C_4$ are integration constants. From (\ref{steph77}), by writing the radicand as $\left[5-9h_{x\,in}(1+\omega)\right]^2+12h_{x\,in}(1+\omega)$, one observes that, for any $\omega>-1$, $\gamma_{+}>0$ and $\gamma_{-}<0$. Moreover, in this context of asymptotic analysis, we observe from (\ref{steph63}) that for small $t$, the torsion scalar $T$ is large, while for large $t$, the torsion is small. Thus, for small $t$ with $h_{x\,in}>1$, corresponding to the inflation, the algebraic expression of $f(T)$ is given by
\begin{eqnarray}
f(T)=C_3T^{\gamma_{+}}\,\,\,.
\end{eqnarray}
Since $h_x(t)$ reduces to $h_{x\,out}$ in the late universe, the model corresponding to the late accelerated universe can be obtained by replacing $h_{x\,in}$ by $h_{x\,out}$. Precisely, for large $t$, the torsion scalar is small, and for $h_{x\,out}>1$, the dominate term in (\ref{steph77}), corresponding to the model of late time universe, is
\begin{eqnarray}
f(T)=C_4T^{\gamma_{-}'}\,\,\,,\;\;\gamma_{\pm}'=\frac{5-9h_{x\,out}(1+\omega)\pm\sqrt{25-78h_{x\,out}(1+\omega)+81h^2_{x\,out}(1+\omega)^2}}{4}\,\,.
\end{eqnarray}
This model is equivalent to the teleparallel gravity for $C_4=1$ and $h_{x\,out}=2/(5+5\omega)$. It is easy to see from this that, for any ordinary matter, i.e., $\omega>0$, one gets $h_{x\,out}<1$, meaning that the teleparallel gravity without cosmological constant cannot provide the late acceleration of the universe (remembering that the acceleration is guaranteed for $h_{x\,out}>1$, and $0<h_{x\,out}<1$ characterising a decelerated expanding universe). Thus, the contribution of the $f(T)$ terms plays the role of the dark energy.\par
Looking for the expression of $f(T)$ for large cosmic time $t$, i.e., the expression (\ref{steph77}) (replacing $h_{x\,in}$ by $h_{x\,out}$), a similarity can be observed with a result of Bamba et al in \cite{bambamyr}. In this work, they undertook $f(T)$ theory in the FLRW metric, first assuming a power-law expression for $f(T)$ in the Eq.~(4.6) for investigating what type of finite future time singularities may appear. Later, they introduced a correction term, still in the form of power-law, their Eq.~(4.22), in order to analyse the possible avoidance of the singularities. Then, they obtained the global expression (4.23) of \cite{bambamyr}. Note that this expression is equivalent to our Eq.~(\ref{steph77}). Moreover, they shown in their ``TABLE II" that the Big Rip and the Big Freeze can be removed if the product of the exponents is negative. This is exactly our result, since $\gamma_{+}'$ and $\gamma_{-}'$ are always positive and negative, respectively ($\gamma_{\pm}'$ are obtained from $\gamma_{\pm}$ by replacing $h_{x\,in}$ by $h_{x\,out}$). This means that for both Bianchi type-I and KS, if the expansion of the universe occurs with the same rate in all direction, models that can realize the late time accelerated expansion of the universe, and able to prevent the Big Rip and the Big Freeze can be reconstructed.

\subsubsection{Using an auxiliary scalar field}\label{numero2}
In this subsection, we would like to use the reconstruction scheme, for which an auxiliary scalar field is introduced in the action of the theory. By this way, the functional form of $f(T)$ can be found through two other scalar functions $P$ and $Q$
(for more clarification on the method, see the case of $f(R)$ gravity \cite{odintsov-asf} and other of $f(T)$ gravity \cite{daouda3}).
Here, we return to the general system (\ref{steph68})-(\ref{steph72}), where we introduce the auxiliary scalar field $\phi$, by defining the algebraic function $f(T)$ by
\begin{eqnarray}
f(T)=P(\phi)T+Q(\phi)\,\,\,. \label{steph78}
\end{eqnarray}
By using (\ref{steph78}) and varying the action (\ref{action}) with respect to the scalar field $\phi$, one gets
\begin{eqnarray}
P'(\phi)T+Q'(\phi)=0\,\,\,,
\end{eqnarray}
which may be solved with respect to $\phi$, yielding  $\phi=\phi(T)$. Here, the prime (') denotes the derivative with respect to $\phi$. By using (\ref{steph78}), one obtains
\begin{eqnarray}
f_T(T)=P(\phi(T)) \,\,\,,\quad f_{TT}(T)=P_T(\phi(T))\,\,\,. \label{steph80}
\end{eqnarray}
Making use of (\ref{steph80}), one can rewrite the system (\ref{steph68})-(\ref{steph71}) as 
\begin{eqnarray}
16\pi\rho&=&P(\phi)T+Q(\phi)-2TP(\phi)\,\,\,,\label{steph81}\\
-16\pi p_x&=&P(\phi)T+Q(\phi)+2P(\phi)\left[\frac{(h_y+h_z)^2-(h_y+h_z)-h_yh_z}{\Psi_0(T)}-\frac{T}{2}\right]\nonumber\\&+&4\left(\frac{h_y+h_z}{\Psi_0(T)}\right){T}P_T(\phi)\,\,,\label{steph82}
\end{eqnarray}
\begin{eqnarray}
-16\pi p_y&=&P(\phi)T+Q(\phi)+2P(\phi)\left[\frac{(h_x+h_z)^2-(h_x+h_z)-h_xh_z}{\Psi_0(T)}-\frac{T}{2}\right]\nonumber\\
&+&4\left(\frac{h_x+h_z}{\Psi_0(T)}\right){T}P_T(\phi)\,\,,\label{steph83}\\
-16\pi p_z&=&P(\phi)T+Q(\phi)+2P(\phi)\left[\frac{(h_x+h_y)^2-(h_x+h_y)-h_xh_y}{\Psi_0(T)}-\frac{T}{2}\right]\nonumber\\
&+&4\left(\frac{h_x+h_y}{\Psi_0(T)}\right){T}P_T(\phi)\,\,\label{steph84} \,\,.
\end{eqnarray}
Let us consider that $p_y=p_z$. Then, by equating (\ref{steph83}) with (\ref{steph84}), one gets
\begin{eqnarray}
TP_T(\phi)=-\frac{1}{2}P(\phi)\left(h_x+h_y+h_z-1\right)\,\,\,.\label{steph85}
\end{eqnarray}
Using (\ref{steph85}), the system (\ref{steph81})-(\ref{steph84}) reduces to
\begin{eqnarray}
16\pi\rho&=&Q(\phi)-TP(\phi)\,\,\,,\label{steph86}\\
-16\pi p_x&=&Q(\phi)+TP(\phi)\,\,\,,\label{steph87}\\
p_x&=& p_y=p_z \,,\,\omega_x=\omega_y=\omega_z\label{steph88}\,.
\end{eqnarray} 
By using (\ref{steph86}), one can determine $Q(\phi)$ as
\begin{eqnarray}
Q(\phi)=16\pi~\rho+TP(\phi)\,\,\,,\label{steph90}
\end{eqnarray}
which, substituted in (\ref{steph87}) and (\ref{steph88}), yields
\begin{eqnarray}
-16\pi(\omega_x+1)\rho=2TP(\phi)\label{steph91}\,.
\end{eqnarray}
As one can redefine the scalar field properly, we may choose $\phi=t$, then $P(\phi)=P(t)=\bar{P}(T)$. Note that in the case where $\omega_x=-1$, one gets $\bar{P}(T)=0$, and the algebraic function $f(T)=Q(t)=16\pi\rho=\bar{Q}(T)$. But this requires to have the complete expression of $\rho$ depending on $T$, i.e., solving the equation of continuity. Let us consider in general the case where $\omega_x\neq-1$ and try to solve the equation of continuity, which will help us to determine  $\bar{P}(T)$ and $\bar{Q}(T)$, leading to the reconstruction of the algebraic function $f(T)$. We choose the case of KS geometry ($h_y=h_z$), where the equation of continuity is written as
\begin{eqnarray}
\dot{\rho}+\rho(1+\omega_x)(H_x+2H_y)=0\,\,\,,
\end{eqnarray}
which can be solved, giving
\begin{eqnarray}
\rho(t)&=&C_5\exp{\left[G_1(t)+G_2(t)\right]}\label{steph93}\,\,\,,\\
G_1(t)&=&\frac{(1+\omega_x)(h_{x\,in}+2h_{y\,in})}{t}\,\,\,,\\
G_2(t)&=&\sqrt{q}\left[(1+\omega_x)(h_{x\,in}-h_{x\,out}+2h_{y\,in}-2h_{y\,out})\right]
\arctan{\left(\sqrt{q}t\right)}\,\,\,,
\end{eqnarray}
where $C_5$ is an integration constant.
Making use of (\ref{steph67}), one can cast $\rho(t)$ in terms of $T$. Also, from Eq.~(\ref{steph91}), one obtains $P(t)$ in functions of $T$, and from (\ref{steph90}), one gets $Q(t)$ in terms of $T$. Therefore, we can reconstruct the algebraic function  $f(T)$, given in  (\ref{steph78}), as
\begin{eqnarray}
f(T)&=&-16\pi C_3 \omega_x \exp\left[\frac{g_1}{\sqrt{\Psi_0 (T)}}+g_2 \arctan\left(\sqrt{q\Psi_0 (T)}\right)\right]\label{steph103}\,,
\end{eqnarray}
where $g_1=(1+\omega_x)(h_{x\,in}+2h_{y\,in})$ and $g_2=\sqrt{q}(1+\omega_x)(h_{x\,in}-h_{x\,out}+2h_{y\,in}-2h_{y\,out})$.
\par
In principle, with (\ref{steph103}), some cosmological models can now be reproduced. Let us focus our attention to the  early universe (which may be characterized by the inflation), and the late time universe (characterized by its accelerated expansion) in the three spatial direction. \par
At early time, i.e., for small $t$  (or  large $T$), $\Psi_0(T)$ is very small, and the corresponding algebraic function is
\begin{eqnarray}
f(T)=-16\pi C_3\omega_x\exp{\left[\frac{g_1}{\sqrt{\Psi_0 (T)}}\right]}\,\,\,.
\end{eqnarray}
At late time, the time $t$ is large (corresponding to small torsion scalar and large $\Psi_0(T)$), and the algebraic function reads
\begin{eqnarray}
f(T)=-16\pi C_3\omega_x\exp{\left[g_2 \arctan\left(\sqrt{q\Psi_0 (T)}\right)\right]}\,\,\,.
\end{eqnarray}
Thus, we see that models corresponding to the inflation and the late time accelerated universe can be reconstructed within KS metrics where the matter content is partially isotropic ($p_y=p_z$). Since in this work we fall in the situation where $\omega_x=\omega_y=\omega_y$ one could use the WMAP data and try to check if they fit with this anisotropic model. Because of having ultimately $p_x=p_y=p_z$ and $h_y=h_z$, one can just use the first two equations of motion of KS case, i.e., (\ref{densityks})-(\ref{radialpressureks}). In order to put out the contribution of the modified part of the algebraic function $f(T)$, we cast it into the form $f(T)=T+j(T)$, as the teleraparallel term plus the modified additive algebraic function $j(T)$. Thus, the equations (\ref{densityks}) and (\ref{tangentialpressureks}) become
\begin{eqnarray}
8\pi \rho_{eff}&=& \left(\frac{\dot{B}}{B}\right)^2+2\frac{\dot{A}\dot{B}}{AB}\,\,,\label{effectivedensity}\\
-8\pi p_{x\,eff}&=& 2 \frac{\ddot{B}}{B}+\left(\frac{\dot{B}}{B}\right)^2\,\,,\label{effectivepressure}
\end{eqnarray}
where $\rho_{eff}$ and $p_{x\,eff}$ are the effective energy density and effective pressure, respectively, and defined by
\begin{eqnarray}
\rho_{eff}&=&\rho-\frac{1}{16\pi}\left\{j+4j_T\left[ \left(\frac{\dot{B}}{B}\right)^2+2\frac{\dot{A}\dot{B}}{AB}\right]\right\}\,\,\,,\\
p_{x\,eff}&=&p_{x}+\frac{1}{16\pi}\left\{j+4j_T\left[\frac{\ddot{B}}{B}+\left(\frac{\dot{B}}{B}\right)^2+\frac{\dot{A}\dot{B}}{AB}\right]+4\frac{\dot{B}}{B}\dot{T}j_{TT}\right\}\,\,\,.
\end{eqnarray}
By dividing (\ref{effectivepressure}) by (\ref{effectivedensity}), using $\omega_{eff}=p_{x\,eff}/\rho_{eff}$, one gets
\begin{eqnarray}
\omega_{eff}=\frac{2-3h_y}{h_y+2h_x}\label{steph104}\,.
\end{eqnarray}
Since the observational data from the WMAP project do not depend on the spatial direction, i.e., are not based on an anisotropic geometry, we have to impose $h_x=h_y$ in (\ref{steph104}), for getting a suitable cosmological feature. Thus, assuming that the accelerated expansion of the universe is realized with the same rate in all directions, $h_x=h_y=h_z>1$, and that the universe is essentially filled  by the dark energy, where we can neglect the usual matter content such that $\omega_{eff}\sim \omega_{DE}$. Therefore, using (\ref{steph104}), we have the possibility to regain the well known range of values  allowed by the current 7-year WMAP data for the parameter of the equation of state of the  dark energy, $\omega_{DE}= -1.1 \pm 0.14$  WMAP \cite{wmap}.

\subsection{On Bianchi type-III solutions}

In this section, we propose to present some comments on Bianchi type-III solutions. This case is quite exceptional since we do not have the freedom of making cosmological reconstruction as in the case of Bianchi type-I and KS, due to the constraints equations 
(\ref{constraint1}) and (\ref{constraint2}). \par
From (\ref{constraint2}), since the parameter is different from zero and the algebraic function cannot be a constant, one gets
\begin{eqnarray}
\frac{\dot{A}}{A}=\frac{\dot{B}}{B}\,\,\,,
\end{eqnarray}
which, injected in (\ref{constraint1}) leads to
\begin{eqnarray}
\dot{T}f_{TT}=0\,\,\,,\label{constraint3}
\end{eqnarray}
meaning that one has $\dot{T}=0$ or $f_{TT}=0$. The first case, $\dot{T}=0$ implies that one has a constant torsion scalar, i.e.,
\begin{eqnarray}
\frac{\dot{A}^2}{A^2}+2\frac{\dot{A}\dot{C}}{AC}=K\,\,,\label{steph111}
\end{eqnarray}
where $K$ is positive constant. Let us consider $A=C^n$, with $n$ bigger than zero or less  than $-2$. Thus, Eq.~(\ref{steph111}) can be solved yielding
\begin{eqnarray}
C(t)=C_0'\exp{\left(\pm\sqrt{\frac{K}{n(n+2)}}\;t\;\right)}\,\,\,,\label{soluC}
\end{eqnarray}
leading to 
\begin{eqnarray}
A(t)=B(t)=(C_0')^n\exp{\left(\pm n\sqrt{\frac{K}{n(n+2)}}\;t\;\right)}\,\,\,,
\end{eqnarray}
where $C_0'$ is a positive constant. It is important to note that in order to guarantee the expansion of the universe, one need to have
\begin{eqnarray}\label{vincent18}
A(t)=B(t)=\left\{\begin{array}{ll}
(C_0')^n\exp{\left(- n\sqrt{\frac{K}{n(n+2)}}\;t\;\right)}\,\,\,,\quad \mbox{for}\,\,\, n<-2 \\
(C_0')^n\exp{\left( n\sqrt{\frac{K}{n(n+2)}}\;t\;\right)}\,\,\,, \;\;\;\quad \mbox{for}\,\,\, n>0\,\,\,.\end{array}\right.
\end{eqnarray}

In this case, we see that the rate of expansion is constant for the three spatial direction: this is the de Sitter universe.\par
Now we can perform the reconstruction of the algebraic function $f(T)$.  One can cast Eqs.~(\ref{densitytype3})-(\ref{tangentialpressure2type3}) in the following system 
\begin{eqnarray}
16\pi\rho&=&f+4f_T\left(K-\frac{\alpha^2}{2A^2}\right)\,\,\,,\label{jesuis}\\
-16\pi\omega_x\rho&=&f+2Kf_T\left[\frac{2n^2+3n+1}{n(n+2)}\right]\,\,\,,\label{tues}\\
-16\pi\omega_z\rho&=&f+4Kf_T\left(\frac{2n+1}{n+2}\right)-2\frac{\alpha^2}{A^2}f_T\;\;,\label{ilest}\\
p_x&=&p_y\,\,\,.
\end{eqnarray}
By combining (\ref{jesuis}) with (\ref{ilest}), one can eliminate the term containing $\alpha$, obtaining
\begin{eqnarray}
-16\pi(\omega_z+1)\rho=\frac{4K(n-1)}{n+2}f_T \,\,.\label{noussommes}
\end{eqnarray}
The energy density $\rho$ can be eliminated by combining (\ref{tues}) with (\ref{noussommes}) yielding the following differential equation
\begin{eqnarray}
2K\left[2n(n-1)\omega_x-(\omega_z+1)(2n^2+3n+1)\right]f_T-n(n+2)(\omega_z+1)f=0\,\,\,,
\end{eqnarray}
whose general solution is 
\begin{eqnarray}
f(T)&=&C_6\exp{\left[R(n)T\right]}\,\,\,,\\
R(n)&=&\frac{n(n+2)(\omega_z+1)}{2K\left[2n(n-1)\omega_x-(\omega_z+1)(2n^2+3n+1)\right]}\,\,,\nonumber
\end{eqnarray}
where $C_6$ is an integration constant. Note that for $n=1$ and $\omega_x=\omega_z$, Eq.~(\ref{D10}) is recovered.
\par
The second case from (\ref{constraint3}), $f_{TT}=0$, implies that $f_{T}$ is constant, that we choose to be minus two times the cosmological constant $\Lambda$, then, $f(T)$ is written as
\begin{eqnarray}
f(T)=T-2\Lambda\,\,\,,
\end{eqnarray}
which is the teleparallel gravity with cosmological constant.


\section{Conclusion} \label{conclusions}

Along the paper, the Bianchi type-I, Kantowski-Sachs, and Bianchi type-III metrics have been studied in the context of $f(T)$ gravities. Particularly, we have shown the reconstruction of some important cosmological solutions, obtaining the corresponding $f(T)$ action. We have assumed initially a particular choice of coordinates and tetrads, specifically cartesian coordinates and a diagonal set of tetrads have been imposed in order to avoid the well known constraint $f_{TT}=0$, which reduces trivially to the action of teleparallel gravity (see Ref.~\cite{Tamanini:2012hg}).\par
Then, several important cosmological solutions have been considered. In particular, dS solutions, where the scale factor is an exponential function of the cosmic time, has been considered for Bianchi type-I and Kantowski-Sachs metrics by imposing a particular exponential expansion in each direction of the space.  We have shown that the only possible solution turns out to the FLRW metric, such that no possible dS anisotropic evolution can be found in $f(T)$, unless one considers an anisotropic fluid. Nevertheless, in the case of power law solutions, we have found that in the presence of a perfect isotropic fluid, an anisotropic cosmological evolution can be found for a particular choice of the action $f(T)$, while in vacuum the action reduces to FLRW metric.\par
Moreover, we have extended the cosmological reconstruction scheme to a general exponential solutions, from which the above de Sitter law and power law solutions are particular cases. We have assumed an adiabatic approximation for the expansion in each spatial direction. We undertook two cases, a special case and a second where an auxiliary field is used. In the both cases, we shown that the models can realize the early accelerated universe, characterized by the inflation, and the late time acceleration of our current universe. In the special case, the model presents an interesting aspect because it ensures the avoidance of the Big Rip and the Big Freeze. In the case where the auxiliary field is used, the model corresponding to the late time accelerated universe fits with the 7-year WMAP data, confirming the consistency of the result.
\par
The Bianchi type-III case  presents some constraints from which only two forms of the algebraic function $f(T)$ can be obtained. The first is the well known teleparallel gravity with cosmological constant, and the second is a de Sitter type solution.

\vspace{0,25cm}
{\bf Acknowlegments} The authors thank Prof. J. D. Barrow for useful discussions. MER thanks  UFES for the hospitality during the development of this work. MJSH thanks CNPq/FAPES for financial support. DSG acknowledges support from a postdoctoral contract from the University of the Basque Country (UPV/EHU) under the program ``Specialization of research staff'', and support from the research project FIS2010-15640, and also by the Basque Government through the special research action KATEA and UPV/EHU under program UFI 11/55.



\begin{thebibliography}{10}
\bibitem{riess}A. G. Riess et al. [Supernova Search Team Collaboration], Astron. J. {\bf 116}: 1009 (1998).

\bibitem{copeland}Edmund J. Copeland, M. Sami and Shinji Tsujikawa, Int.J.Mod.Phys. D {\bf 15} (2006) 1753-1936, arXiv:hep-th/0603057v3;
  T.~Padmanabhan,
  AIP Conf.\ Proc.\  {\bf 843}, 111 (2006)
  [astro-ph/0602117];
  J.~Frieman, M.~Turner and D.~Huterer,
  Ann.\ Rev.\ Astron.\ Astrophys.\  {\bf 46}, 385 (2008)
  [arXiv:0803.0982 [astro-ph]];
  P.~J.~E.~Peebles and B.~Ratra,
  Rev.\ Mod.\ Phys.\  {\bf 75}, 559 (2003)
  [astro-ph/0207347].


\bibitem{reviews}
S. Nojiri and S. D. Odintsov, Phys. Rept. {\bf 505}, 59-144 (2011); S. Nojiri and S.D. Odintsov, ECONF
C 0602061:06, (2006); Int. J. Geom. Meth. Mod. Phys. {\bf 4} 115-146, (2007);
S. Capozziello and V. Faraoni, Beyond Einstein Gravity, A Survey of Gravitational Theories for Cosmology and Astrophysics, Series: Fundamental Theories of Physics, Vol. 170, Springer, New York (2011); 
  S.~Capozziello and M.~De Laurentis,
  Phys.\ Rept.\  {\bf 509}, 167 (2011)
  [arXiv:1108.6266 [gr-qc]];
  T.~Clifton, P.~G.~Ferreira, A.~Padilla and C.~Skordis,
  Phys.\ Rept.\  {\bf 513}, 1 (2012)
  [arXiv:1106.2476 [astro-ph.CO]].
  A.~de la Cruz-Dombriz and D.~Saez-Gomez,
  arXiv:1207.2663 [gr-qc].
  
  \bibitem{Bamba:2012cp} 
  K.~Bamba, S.~Capozziello, S.~'i.~Nojiri and S.~D.~Odintsov,
  arXiv:1205.3421 [gr-qc].
  

\bibitem{pereira1}R. Aldrovandi and J. G. Pereira, TELEPARALLEL GRAVITY, in http://www.ift.unesp.br/ users/jpereira/tele.pdf.

\bibitem{ferraro1}R. Ferraro and F. Fiorini, Phys. Lett. B {\bf 702} (2011) 75 [arXiv:1103.0824 [gr-qc]].

\bibitem{ferraro2}R. Ferraro and F. Fiorini, Int. J. Mod. Phys. Conf. Ser. {\bf 3}  227 (2011), [arXiv:1106.6349 [gr-qc]].

\bibitem{ferraro3} R.~Ferraro and F.~Fiorini, Phys.\ Rev.\ D {\bf 75} 084031 (2007), gr-qc/0610067.

\bibitem{ferraro4}R.~Ferraro and F.~Fiorini, Phys.\ Rev.\ D {\bf 78} 124019 (2008), arXiv:0812.1981 [gr-qc].

\bibitem{gabriel}G.~R.~Bengochea and R.~Ferraro, Phys.\ Rev.\ D {\bf 79} 124019 (2009), arXiv:0812.1205 [astro-ph].

\bibitem{Linder:2010py} 
  E.~V.~Linder,
  Phys.\ Rev.\ D {\bf 81}, 127301 (2010)
  [Erratum-ibid.\ D {\bf 82}, 109902 (2010)]
  [arXiv:1005.3039 [astro-ph.CO]].

\bibitem{wu}Puxun Wu and Hongwei Yu, 	Phys.\ Lett.\ B {\bf 693} 415-420 (2010), arXiv:1006.0674v5 [gr-qc].  

\bibitem{ratbay}Ratbay Myrzakulov, Eur.\ Phys.\ J.\ C {\bf 71}, 1752 (2011), arXiv:1006.1120v1 [gr-qc].

\bibitem{barrow}L.~Baojiu, T.~P.~Sotiriou, and J.D. Barrow, Phys. Rev. D {\bf 83}, 064035 (2011); Phys.Rev.D {\bf 83}:104030 (2011).

\bibitem{daouda1}M. Hamani Daouda, Manuel E. Rodrigues, M. J. S. Houndjo, Eur. Phys. J. C. {\bf 71} 1817 (2011), arXiv:1108.2920v4 [astro-ph.CO]; Euro. Phys. J. C {\bf 72} 1890 (2012), arXiv:1109.0528v4 [physics.gen-ph].

\bibitem{daouda3}M.~H.~Daouda, M.~E.~Rodrigues and M.~J.~S.~Houndjo, Phys.\ Lett.\ B {\bf 715}, 241 (2012), arXiv:1202.1147v2 [gr-qc].

\bibitem{boehmer}Christian G. Boehmer, Tiberiu Harko and Francisco S. N. Lobo, Phys.\ Rev.\ D{\bf 85} 044033 (2012), arXiv:1110.5756v2 [gr-qc]. 

\bibitem{x}Jie Yang, Yun-Liang Li, Yuan Zhong and Yang Li, arXiv:1202.0129v1 [hep-th]; K. Karami and A. Abdolmaleki, arXiv:1201.2511v1 [gr-qc]; K. Atazadeh and F. Darabi, arXiv:1112.2824v1 [physics.gen-ph]; Hao Wei, Xiao-Jiao Guo and Long-Fei Wang, Phys.Lett.B {\bf 707}:298-304 (2012); K. Karami, A. Abdolmaleki, arXiv:1111.7269v1 [gr-qc]; P.A. Gonzalez, Emmanuel N. Saridakis  and Yerko Vasquez, arXiv:1110.4024v1 [gr-qc]; S. Capozziello, V. F. Cardone, H. Farajollahi and A. Ravanpak, Phys.Rev.D {\bf 84}:043527 (2011); Rong-Xin Miao, Miao Li and Yan-Gang Miao, arXiv:1107.0515v3 [hep-th]; Xin-he Meng and Ying-bin Wang, Eur.Phys.J. C {\bf 71}:  1755 (2011); Hao Wei, Xiao-Peng Ma and Hao-Yu Qi, Phys.Lett.B {\bf 703}:74-80 (2011); Miao Li, Rong-Xin Miao and Yan-Gang Miao, JHEP {\bf 1107}:108 (2011); Surajit Chattopadhyay and Ujjal Debnath, Int.J.Mod.Phys.D {\bf 20}:1135-1152 (2011); Piyali Bagchi Khatua, Shuvendu Chakraborty and Ujjal Debnath, arXiv:1105.3393v1 [physics.gen-ph]; M. R. Setare and M. J. S. Houndjo, arXiv:1203.1315 [gr-qc];  
Yi-Fu Cai, Shih-Hung Chen, James B. Dent, Sourish Dutta and Emmanuel N. Saridakis, Class. Quantum Grav. {\bf 28}:  215011 (2011); Rong-Jia Yang, Europhys.Lett. {\bf 93}:60001 (2011); Christian G. Boehmer, Atifah Mussa and Nicola Tamanini, Class.Quant.Grav. {\bf 28}: 245020 (2011); P. A. Gonzalez, Emmanuel N. Saridakis and Yerko Vasquez, JHEP 1207 (2012) 053, arXiv:1110.4024v2 [gr-qc]; 
  A.~Behboodi, S.~Akhshabi and K.~Nozari,
  arXiv:1205.4570 [gr-qc];
Di Liu, Puxun Wu and Hongwei Yu, arXiv:1203.2016v1 [gr-qc].
\bibitem{cemsinan}Cemsinan Deliduman and Baris Yapiskan, arXiv:1103.2225 [gr-qc].

\bibitem{houndjo} M. J. S. Houndjo, D. Momeni and R. Myrzakulov, arXiv:1206.3938v1 [physics.gen-ph].

\bibitem{daouda2}M. Hamani Daouda, Manuel E. Rodrigues and M. J. S. Houndjo, arXiv:1205.0565v1 [gr-qc].

\bibitem{sharif}M. Sharif and Shamaila Rani, Mod. Phys. Lett. A26 (2011)1657-1671, arXiv:1105.6228v1 [gr-qc].

\bibitem{li}L.~Baojiu, Thomas P.~Sotiriou, John D. Barrow, Phys.\ Rev.\ D {\bf 83} 104017 (2011);  Shih-Hung Chen, J. B. Dent, S. Dutta and E. N. Saridakis, Phys.\ Rev.\ D {\bf 83} 023508 (2011).


 \bibitem{LittleRip}
  P.~H.~Frampton, K.~J.~Ludwick and R.~J.~Scherrer,
  Phys.\ Rev.\ D {\bf 84}, 063003 (2011)
  arXiv:1106.4996 [astro-ph.CO].

 \bibitem{Nojiri:2011kd} 
  S.~Nojiri, S.~D.~Odintsov and D.~S\'aez-G\'omez, AIP Conf.\ Proc.\  {\bf 1458}, 207 (2011), 
  arXiv:1108.0767 [hep-th];
  A.~J.~L\'opez-Revelles, R.~Myrzakulov and D.~S\'aez-G\'omez,
  Phys.\ Rev.\ D {\bf 85}, 103521 (2012)
  arXiv:1201.5647 [gr-qc].

\bibitem{bambamyr} K. Bamba, R. myrzakulov, S. Nojiri and S. D. Odintsov, Phys. Rev. D {\bf 85}, 104036 (2012). arXiv: 1202.4057[gr.qc].


  \bibitem{SaezGomez:2012ek} 
  D.~S\'aez-G\'omez,
  arXiv:1207.5472 [gr-qc].

\bibitem{barrow3}J. Yearsley and J. D. Barrow, Class. Quant. Grav. {\bf 13}: 2693 (1996).

\bibitem{stability} S. Hervik, D. F. Mota and M. Thorsrud, JHEP {bf 11} (2011) 146; J. D. Barrow and S. Hervik, Phys. Rev. D {\bf 73}: 023007  (2006); Chiang-Mei Chen and W. F. Kao, Phys. Rev. D {\bf 64}: 124019  (2001); J. D. Barrow and K. Yamamoto, Phys. Rev. D {\bf 85}: 083505 (2012); W. F. Kao and Ing-Chen Lin, Phys. Rev. D {\bf 83}: 063004 (2011); JCAP {\bf 01} (2009) 022; G. Leon and E. N. Saridakis, Class.Quant.Grav. {\bf 28}: 065008 (2011).   

\bibitem{STFR}
S. Nojiri and S. D. Odintsov, Phys. Rev. D {\bf 68}, 123512 (2003), [arXiv:hep-th/0307288].

\bibitem{wmap}D. N. Spergel et al., Astrophys. J. Suppl. {\bf 170}, 377 (2007); L. Page et al., ibid., 335 (2007); C. L. Bennett et al., ibid. {\bf 148}, 1 (2003); G. Hinshaw et al., ibid. {\bf 148}, 135 (2003); D. N. Spergel et al., ibid. {\bf 148}, 175 (2003); G. Hinshaw et al., ibid. {\bf 170}, 288 (2007). E. Komatsu et al., ``Seven-Year Wilkinson Microwave Anisotropy Probe (WMAP) Observations: Cosmological Interpretation",     Astrophys. J. Suppl. {\bf 192}, 18 (2011). arXiv:1001.4538 [astro-ph.CO]. 

\bibitem{campanelli1}L. Campanelli, P. Cea and L. Tedesco, Phys. Rev. Lett. {\bf 97}: 131302 (2006); Erratum-ibid. {\bf 97}: 209903 (2006); Phys. Rev. D {\bf 76}: 063007 (2007).

\bibitem{lqc}A. Corichi and E. Montoya, Phys. Rev. D {\bf 85}: 104052  (2012); P. Singh, Phys. Rev. D {\bf 85}: 104011 (2012); B. Gupt and P. Singh, Phys. Rev. D {\bf 85}: 024034 (2012); K. Fujio and T. Futamase, Phys. Rev. D {\bf 85}: 124002 (2012); E. Wilson-Ewing, Phys. Rev. D {\bf 82}: 043508 (2010); A. Ashtekar and E. Wilson-Ewing, Phys. Rev. D {\bf 80}: 123532 (2009); L. Szulc, Phys. Rev. D {\bf 78}: 064035 (2008); Dah-Wei Chiou, Phys. Rev. D {\bf 76}: 124037 (2007); Phys. Rev. D {\bf 75}: 024029 (2007); P. Dzierzak and W. Piechocki, Phys. Rev. D {\bf 80}: 124033 (2009).    

\bibitem{campanelli2}L. Campanelli, Phys. Rev. D {\bf 84}: 123521 (2011).

\bibitem{campanelli3}L. Campanelli, P. Cea, G.L. Fogli, L. Tedesco, Mod.Phys.Lett. A {\bf 26}: 1169-1181 (2011).

\bibitem{fontanini}Michele Fontanini, Mark Trodden and Eric J. West, Phys. Rev. D {\bf 80}: 123515 (2009).

\bibitem{barrow2}J. D. Barrow, Phys. Rev. D {\bf 55}: 7451-7460  (1997).

\bibitem{massimo}Massimo Giovannini, Phys. Rev. D {\bf 59}: 123518  (1999).

\bibitem{inflation}B. Himmetoglu, JCAP {\bf 1003}: 023 (2010); Masa-aki Watanabe, S. Kanno and J. Soda, Phys.Rev.Lett. {\bf 102}: 191302 (2009); C. Pitrou, T. S. Pereira and Jean-Philippe Uzan, 	JCAP {\bf 0804}: 004 (2008); B. C. Paul, Phys. Rev. D {\bf 64}: 124001 (2001).   

\bibitem{x-1}Kei Yamamoto, Phys.Rev. D {\bf 85}: 043510 (2012); Masato Minamitsuji, Phys. Rev. D {\bf 85}: 103526 (2012); M. Sharif and S. Waheed, Eur.Phys.J. C {\bf 72}: 1876 (2012); Hyeong-Chan Kim and M. Minamitsuji, JCAP {\bf 1103}: 038  (2011); Phys. Rev. D {\bf 81}: 083517  (2010), Erratum-ibid. D {\bf 82}: 109904 (2010); L. Campanelli, P. Cea, G. L. Fogli and A. Marrone, Phys. Rev. D {\bf 83}: 103503 (2011); Campanelli, P. Cea, G. L. Fogli and L. Tedesco, Int.J.Mod.Phys. D {\bf20}: 1153-1166 (2011);  J. D. Barrow, Phys. Rev. D {\bf 81}: 023513 (2010); Phys.Rev. D {\bf 59}: 043502 (1999); J. Adamek, D. Campo and J. C. Niemeyer, Phys.Rev. D {\bf 82}: 086006  (2010); L. Campanelli, Phys.Rev. D {\bf 80}: 063006 (2009); A. Pontzen, Phys.Rev. D {\bf 79}: 103518  (2009); D. C. Rodrigues, Phys.Rev. D {\bf 77}: 023534 (2008); Phys.Rev. D {\bf 78}: 063013 (2008); T. Koivisto and D. F. Mota, JCAP {\bf 0806}: 018 (2008); A.E. Gumrukcuoglu, L. Kofman and M. Peloso, Phys.Rev. D {\bf 78}: 103525 (2008); E. J. King and P. Coles, Class.Quant.Grav. {\bf 24}: 2061-2072 (2007); M. Cataldo and S. del Campo, Phys.Rev. D {\bf 62}: 023501 (2000); M. Giovannini, Phys.Rev. D {\bf 59}: 123518 (1999).         

\bibitem{can}C. Aktas, S. Aygun and I. Yilmaz, Phys.Lett. B {\bf 707}: 237-242 (2012);   Astrophys.Space Sci. {\bf 332}: 463-471  (2011); G. Leon and E. N. Saridakis, Class.Quant.Grav. {\bf 28}: 065008 (2011);  M. Sharif and M. Zubair, Int.J.Mod.Phys. D {\bf 19}: 1957-1972 (2010); M. Sharif and M. F. Shamir, Gen.Rel.Grav. {\bf 42}: 2643-2655 (2010); M. F. Shamir, Int.J.Theor.Phys. {\bf 50}: 637-643  (2011).   

\bibitem{sharif2}M. Sharif and H. R. Kausar, Phys.Lett. B {\bf 697}: 1-6  (2011); arXiv:1101.3372 [gr-qc]. 

\bibitem{akarsu}O. Akarsu and C. B. Kilinc, Gen.Rel.Grav. {\bf 42}: 763-775 (2010); arXiv:0909.1025 [gr-qc]. 



\bibitem{Nojiri:2009kx} 
  S..~Nojiri, S.~D.~Odintsov and D.~S\'aez-G\'omez,
  Phys.\ Lett.\ B {\bf 681}, 74 (2009)
  [arXiv:0908.1269 [hep-th]];
  N.~Goheer, J.~Larena and P.~K.~S.~Dunsby,
  Phys.\ Rev.\ D {\bf 80}, 061301 (2009)
  [arXiv:0906.3860 [gr-qc]];
  E.~Elizalde, R.~Myrzakulov, V.~V.~Obukhov and D.~S\'aez-G\'omez,
  Class.\ Quant.\ Grav.\  {\bf 27}, 095007 (2010)
  [arXiv:1001.3636 [gr-qc]];
  R.~Myrzakulov, D.~S\'aez-G\'omez and A.~Tureanu,
  Gen.\ Rel.\ Grav.\  {\bf 43}, 1671 (2011)
  [arXiv:1009.0902 [gr-qc]].



\bibitem{Cognola:2008zp} 
  G.~Cognola, E.~Elizalde, S.~D.~Odintsov, P.~Tretyakov and S.~Zerbini,
  Phys.\ Rev.\ D {\bf 79}, 044001 (2009)
  [arXiv:0810.4989 [gr-qc]].

\bibitem{odintsov-asf} S. Nojiri, S. D. Odintsov, Phys. Rev. D {\bf 74}: 086005 (2006).


\bibitem{Tamanini:2012hg} 
  N.~Tamanini and C.~G.~Boehmer,
  Phys.\ Rev.\ D {\bf 86}, 044009 (2012)
  [arXiv:1204.4593 [gr-qc]].




\end{thebibliography}
\end{document}